\documentclass[pra,aps,twocolumn,amsmath,amssymb,floatfix,superscriptaddress]{revtex4-1}
\usepackage{amsmath}
\usepackage{latexsym}
\usepackage{amssymb}
\usepackage{graphics,epstopdf}
\usepackage[colorlinks=true, citecolor=blue, urlcolor=blue ]{hyperref}
\usepackage{epsf,graphics,graphicx}

\begin{document}

\title{Reentrant localization in a quasiperiodic chain with correlated hopping sequences}

\author{Sourav Karmakar}

\email{karmakarsourav2015@gmail.com}

\affiliation{Physics and Applied Mathematics Unit, Indian Statistical Institute, 203 Barrackpore Trunk Road, Kolkata-700 108, India}

\author{Sudin Ganguly}

\email[corresponding author:] {sudinganguly@gmail.com}

\affiliation{Department of Physics, School of Applied Sciences, University of Science and Technology Meghalaya, Ri-Bhoi-793 101, India}

\author{Santanu K. Maiti}

\email{santanu.maiti@isical.ac.in}

\affiliation{Physics and Applied Mathematics Unit, Indian Statistical Institute, 203 Barrackpore Trunk Road, Kolkata-700 108, India}

\begin{abstract}

Quasiperiodic systems are known to exhibit localization transitions in low dimensions, wherein all electronic states become localized beyond a critical disorder strength. Interestingly, recent studies have uncovered a reentrant localization (RL) phenomenon: upon further increasing the quasiperiodic modulation strength beyond the localization threshold, a subset of previously localized states can become delocalized again within a specific parameter window. While RL transitions have been primarily explored in systems with simple periodic modulations, such as dimerized or long-range hopping integrals, the impact of more intricate or correlated hopping structures on RL behavior remains largely elusive. In this work, we investigate the localization behavior in a one-dimensional lattice featuring staggered, correlated on-site potentials following the Aubry-Andr\'{e}-Harper model, along with off-diagonal hopping modulations structured according to quasiperiodic Fibonacci and Bronze Mean sequences. By systematically analyzing the fractal dimension, inverse participation ratio, and normalized participation ratio, we demonstrate the occurrence of RL transitions induced purely by the interplay between quasiperiodic on-site disorder and correlated hopping. We further examine the parameter space to determine the specific regimes that give rise to RL. Our findings highlight the crucial role of underlying structural correlations in governing localization-delocalization transitions in low-dimensional quasiperiodic systems, where the correlated disorder manifests in both diagonal and off-diagonal terms.

\end{abstract}

\maketitle

\section{Introduction}

The phenomenon of localization has been central to condensed matter physics since Anderson's demonstration of the absence of diffusion in random lattices~\cite{anderson1958absence}, with subsequent works reviewing disordered electronic systems~\cite{lee1985disordered}, tracing its historical development~\cite{abrahams201050}, and detailing the theory of Anderson transitions~\cite{evers2008anderson}. It describes the quantum confinement of particles, 
such as electrons or photons, arising from the destructive interference of wave functions in disordered media. In contrast to periodic 
systems, where Bloch waves represent extended eigenstates, disordered systems can give rise to exponentially localized states. According 
to the scaling theory of localization~\cite{abrahams1979scaling}, all single-particle states in one and two dimensions become localized 
even under infinitesimally weak uncorrelated disorder. However, in three-dimensional systems, a mobility edge can emerge, separating localized from extended states~\cite{mott1987mobility}. This enables a disorder-driven metal-insulator transition. Such behavior has been studied in quantum thermoelectrics~\cite{whitney2014most,yamamoto2017thermoelectricity} and in quasiperiodic heat engines~\cite{chiaracane2020quasiperiodic}. These effects are not 
limited to electronic systems but have been observed in photonic and ultra-cold atomic systems as 
well~\cite{segev2013anderson,b-shapiro}.

Quasiperiodic systems occupy an intermediate regime between periodic and random structures, displaying phenomena such as localization-delocalization transitions at finite modulation strengths~\cite{aubry1980analyticity,harper1955general,jitomirskaya1999metal,aulbach2004phase}, fractal energy spectra~\cite{kohmoto1983localization,ostlund1983one,kohmoto1984cantor,kohmoto1987critical}, and critical eigenstates characteristic of quasicrystals~\cite{suck2013quasicrystals,jagannathan2021fibonacci}. Two paradigmatic examples are the Aubry-Andr\'{e}-Harper (AAH) model~\cite{aubry1980analyticity,harper1955general} and the Fibonacci model~\cite{jagannathan2021fibonacci,roche1997electronic,dal2003light,mace2016fractal}, both of which have been extensively studied for their spectral, transport, and optical properties.
The AAH model introduces incommensurate modulations either in the on-site potentials (diagonal) or in the hopping amplitudes (off-diagonal), or both (generalized). 
In its non-interacting diagonal AAH form, it features a self-dual symmetry leading to a simultaneous localization transition of all eigenstates at a 
critical modulation strength. This self-duality facilitates an exact analytical determination of the transition point, offering deep 
insight into the nature of the extended and localized phases.
On the other hand, the Fibonacci model incorporates a binary modulation of parameters following the Fibonacci substitution rule. Its 
spectrum is singular continuous, and the eigenstates are neither fully localized nor extended but are instead critical with fractal
characteristics at all values of the modulation strength~\cite{kohmoto1983localization,ostlund1983one}. Despite these differences, the AAH and Fibonacci models belong to the same 
topological class and can be viewed as two limiting cases of the interpolating Aubry-Andr\'{e}-Fibonacci (IAAF) model~\cite{kraus2012topological}.

Traditionally, it has been believed that once a system undergoes a localization transition, any further increase in disorder would only enhance localization. However, this conventional understanding of localization with increasing disorder has been challenged by several notable studies~\cite{hiramoto-1989,goblot2020emergence,zhai2021cascade,roy2021reentrant}. In a pioneering work, Hiramoto and Kohmoto~\cite{hiramoto-1989} demonstrated that, in a new class of quasiperiodic potentials, {\it the states near the
edge become extended and then localized reentrantly} as the modulation strength increases, a phenomenon known as reentrant localization (RL). A particularly striking example of RL occurs in the interpolating Aubry-Andr\'{e}-Fibonacci (IAAF) model, which exhibits this behavior throughout its parameter space, as confirmed by recent experiments~\cite{goblot2020emergence}. In such systems, the localization-delocalization transition unfolds in a highly nonuniform fashion across the spectrum, leading to the emergence of multiple mobility edges in both Hermitian\cite{goblot2020emergence} and non-Hermitian\cite{zhai2021cascade} systems. More recently, Roy et al.~\cite{roy2021reentrant} reported a similar RL transition, where certain eigenstates unexpectedly reenter a delocalized phase as modulation strength increases beyond the conventional localization threshold. This anomalous behavior has been attributed to the interplay between hopping dimerization and quasiperiodic modulation, which drives the system beyond standard localization scenarios. Following these pioneering studies, several works have examined RL transitions in greater depth. Non-Hermitian effects have been shown to induce RL behavior in quasiperiodic lattices~\cite{wu2021njp} and in dimerized non-Hermitian systems~\cite{jiang2021cpb}. Disorder-related mechanisms include random-dimer perturbations in Su-Schrieffer-Heeger-type models~\cite{zuopra2022} and multiple localization transitions in quasiperiodic chains with onsite or correlated potentials~\cite{padhanprb2022}. Long-range hopping has also been found to influence the fate of RL transitions in dimerized chains~\cite{wangprb2023}, while phase-shifted quasiperiodic lattices can exhibit multiple reentrant transitions~\cite{liarxiv2023}. Our recent study~\cite{ganguly2023spin} further demonstrated that hopping dimerization is not a prerequisite, with spin-dependent RL behavior arising in antiferromagnetic helices under transverse electric fields.

Although numerous studies have explored RL transitions, the influence of hopping modulation on such behavior remains relatively underexplored. Recently, a study demonstrated RL transitions in a quasiperiodic one-dimensional chain with staggered hopping amplitudes modulated in the frequency domain~\cite{adityaprb2023}. In another study, a 1D off-diagonal IAAF model~\cite{tabaprb2024} exhibited multiple RL transitions, where successive localization-delocalization phases emerged as the interpolating parameter was varied. These observations reveal that the influence of hopping modulation on RL remains scarcely addressed. Considering its demonstrated capacity to generate rich and tunable localization-delocalization transitions, further investigation into this direction holds substantial promise for advancing our understanding and control of wave transport in quasiperiodic systems.

In this work, we consider a one-dimensional tight-binding model featuring a staggered on-site potential in the form of the AAH model and introduce hopping modulations governed by the Fibonacci and bronze mean (BM) sequences~\cite{albuquerque2003theory,macia2005role,guopre2007}. The respective inflation rules are $A \rightarrow AB$, $B \rightarrow A$ for the Fibonacci sequence, and $A \rightarrow AAAB$, $B \rightarrow A$ for the BM sequence. The recursive concatenation follows $F_{n+1} = F_n F_{n-1}$ for Fibonacci and $F_{n+1} = F_n F_n F_n F_{n-1}$ for BM. The hopping integral alternates between two values, $t_A$ and $t_B$, following these sequences via the substitutions $A \rightarrow t_A$ and $B \rightarrow t_B$. Building on this site potential pattern and hopping sequences, we investigate how their interplay drives RL transitions.

The rest of the paper is organized as follows. In Sec. II, we present the model Hamiltonian and describe the numerical techniques 
employed to characterize localization. Section III contains our main results, focusing on the emergence of RL transitions. Finally, 
we conclude in Sec. IV.

\section{Model and the method} 

We consider a one-dimensional tight-binding lattice where the on-site potential is governed by a staggered quasiperiodic modulation 
following the AAH model, and the nearest-neighbor hopping integrals are distributed according to either a Fibonacci or a BM quasiperiodic  
\begin{figure}[ht]
\includegraphics[width=0.45\textwidth]{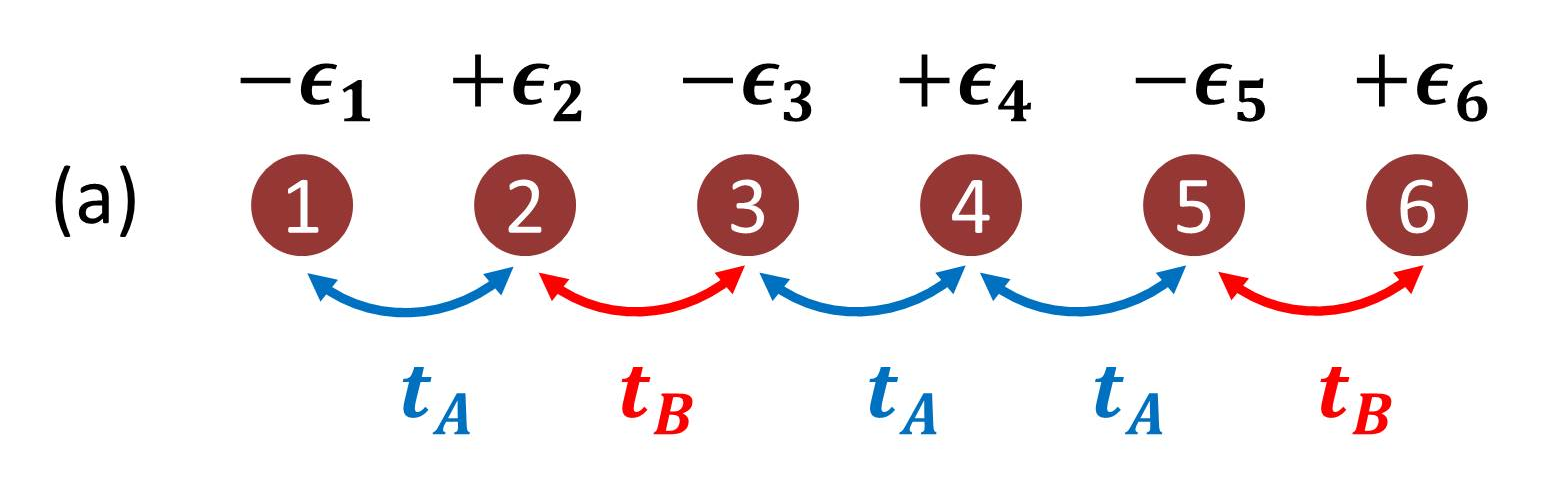}\vskip 0.1 in
\centering \includegraphics[width=0.45\textwidth]{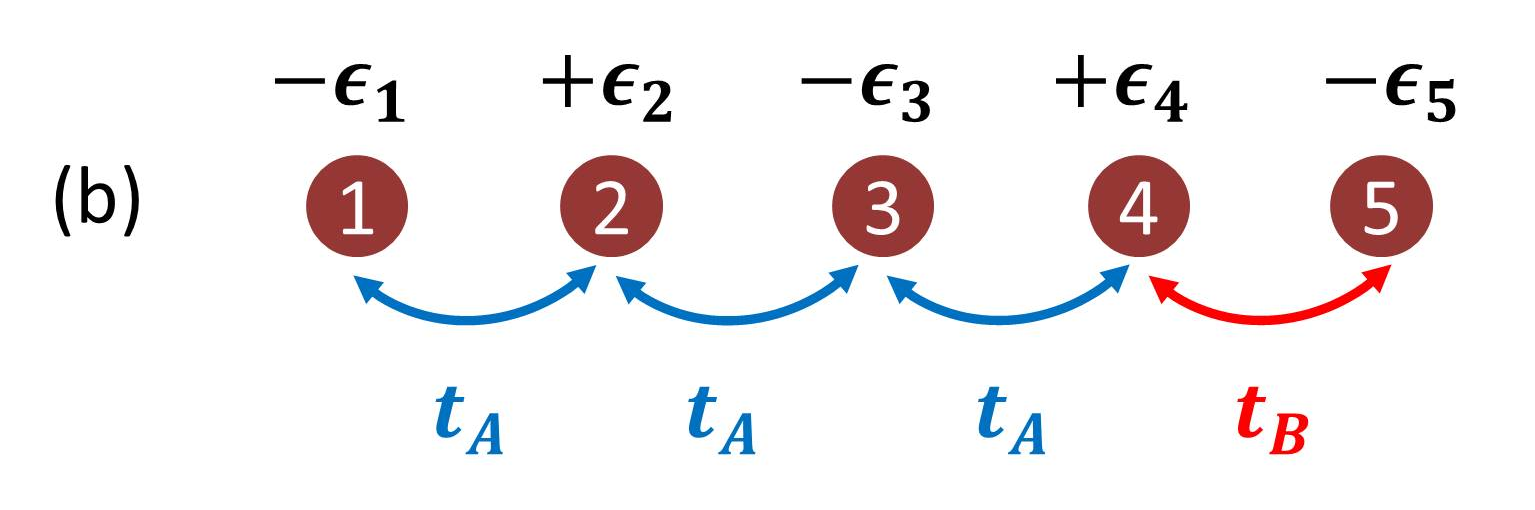}
\caption{(Color online). Quasiperiodic chains with (a) Fibonacci and (b) BM sequences, corresponding to the $4$th and $2$nd generations,
respectively. The on-site energies are staggered, and they are modulated quasiperiodically in the AAH form. The hopping integrals $t_A$ (blue curve) and $t_B$ (red curve) follow the respective sequences generated by the corresponding substitution rules.}
\label{fig:1}
\end{figure}
sequence. Schematic representations of both systems are shown in Figs.~\ref{fig:1}(a) and (b), respectively.

The TB Hamiltonian of the system under open boundary condition can be written as
\begin{equation}
H=\sum_{i} t_i \left(c_i^\dagger c_{i+1} + c_{i+1}^\dagger c_i \right)  + \sum_{i} (-1)^i \epsilon_i c_i^\dagger c_i,
\label{eq:1}
\end{equation}
where $c_i^{\dagger}$ and $c_i$ are the fermionic creation and annihilation operators at site $i$, respectively, and $t_i$ denotes the 
hopping amplitude between the nearest-neighbor sites. The hopping amplitudes take values $t_A$ or $t_B$ depending on the underlying
sequence, Fibonacci or BM~\cite{albuquerque2003theory,macia2005role,guopre2007}. The second term of the Hamiltonian is associated with the staggered on-site energies, where the site energies 
are taken in the AAH form described as~\cite{aubry1980analyticity,harper1955general}
\begin{equation}
\epsilon_i=\lambda \cos(2 \pi b i + \phi).
\label{siteeng}
\end{equation}
Here $\lambda$ denotes the modulation strength, $b$ is an irrational number, and $\phi$ is the AAH phase factor. 
The incommensurate parameter $b$ is chosen according to the ratio between the number of building blocks $t_A$ and $t_B$ in the sequence in the asymptotic limit~\cite{guopre2007}. For the Fibonacci sequence, this ratio tends to the golden mean $(1+\sqrt{5})/2$, while for the BM case, it approaches $(3+\sqrt{13})/2$. Throughout the analysis, we set $\phi=0$ as the phase factor does not have any qualitative effect on the localization properties~\cite{goblot2020emergence}.

To characterize the nature of the eigenstates, whether localized, extended, or critical, we evaluate the inverse participation ratio (IPR) 
and the normalized participation ratio (NPR) for each normalized eigenstate $\psi_n$, defined as follows~\cite{li2017mobility,li2020mobility}.
\begin{eqnarray}
\mathrm{IPR}_n &=& \sum_{i=1}^{N} |\psi_n^i|^4, \nonumber \\
\mathrm{NPR}_n &=& \left(N \sum_{i=1}^{N} |\psi_n^i|^4 \right)^{-1},
\label{eq:2}
\end{eqnarray}
where $N$ is the total number of lattice sites. For an extended state, $\mathrm{IPR}_n\sim 1/N$ drops to zero in the asymptotic limit 
($N\rightarrow \infty$), while for the localized state it reaches to unity. In contrast, $\mathrm{NPR}_n$ approaches unity for extended 
states and vanishes for localized ones.

To examine the global behavior of the spectrum, particularly in identifying mobility edges or intermediate phases, we compute the 
averaged IPR and NPR over a subset of eigenstates $N_L$, and they are defined as~\cite{li2017mobility,li2020mobility},
\begin{eqnarray}
\langle \text{IPR} \rangle &=& \frac{1}{N_L} \sum_{n=1}^{N_L} \text{IPR}_n, \nonumber \\
\langle \text{NPR} \rangle &=& \frac{1}{N_L} \sum_{n=1}^{N_L} \text{NPR}_n.
\label{eq:3}
\end{eqnarray}
In the asymptotic limit, $\langle \text{IPR} \rangle \rightarrow 0$ and $\langle \text{NPR} \rangle \rightarrow 1$ for fully extended 
states, while the opposite holds for a completely localized case. For the critical or mixed phases, both the quantities take intermediate 
values.

Beyond the participation ratios, we also employ the fractal dimension $D_n$ to probe the multifractal nature of 
the eigenstates~\cite{deng2019one,yao2019critical,roy2021fraction}. It is defined through a scaling relation with IPR as
\begin{equation}
D_n = -\lim_{N \to \infty} \frac{\log(\text{IPR}_n)}{\log(N)}.
\label{eq:4}
\end{equation}
This dimension $D_n$ quantifies the extent to which the wave function spreads across the lattice. For extended states, $D_n=1$; for 
localized states, $D_n=0$; and for multifractal (critical) states, $D_n$ lies between $0$ and $1$. Unlike IPR and NPR, this method does 
not require extrapolation or comparison with finite-size benchmarks to distinguish state types~\cite{evers2008anderson}.

\section{Numerical Results and discussion} 
Before we begin our discussion, let us first mention the units and parameters used in this work. All the energies are measured in the unit of eV. Unless mentioned otherwise, the hopping parameters $t_A$ and $t_B$ are fixed at 1$\,$eV and 2.5$\,$eV, respectively. 

\begin{figure}[ht]
\includegraphics[width=0.2\textwidth]{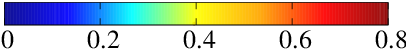}\vskip 0.1 in
\includegraphics[width=0.238\textwidth]{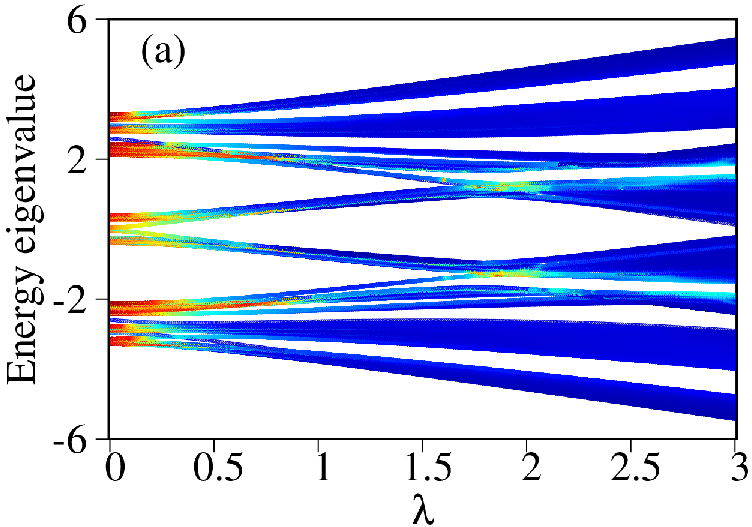}
\hfill
\includegraphics[width=0.238\textwidth]{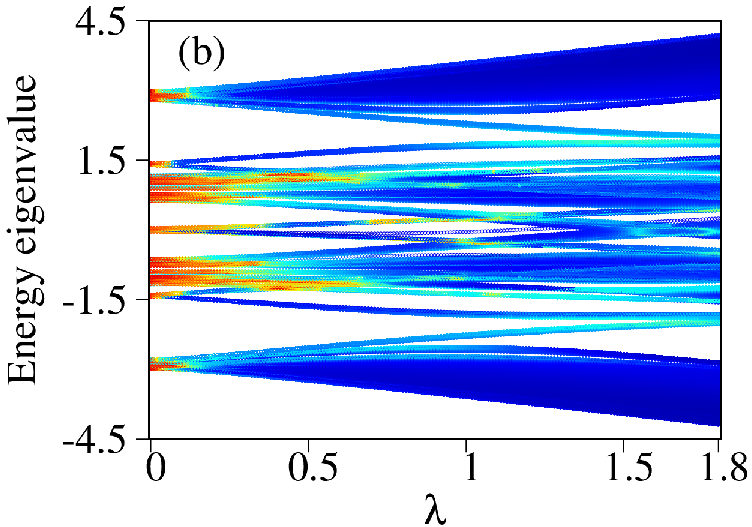}
\vskip 0.1 in
\includegraphics[width=0.238\textwidth]{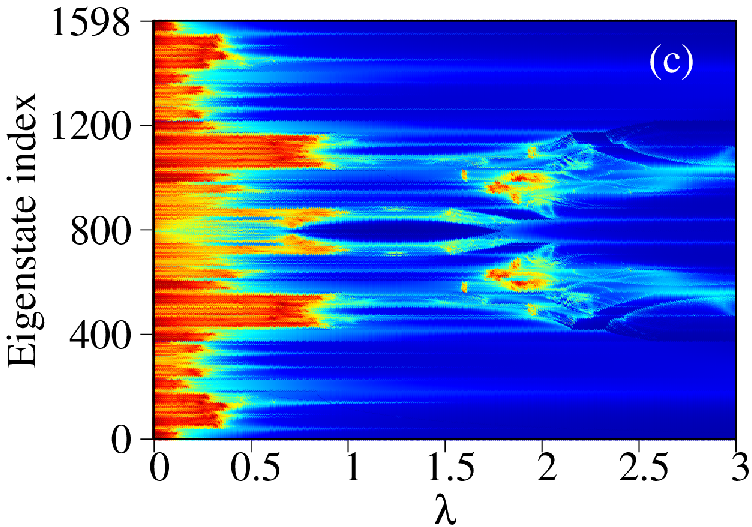}
\hfill
\includegraphics[width=0.238\textwidth]{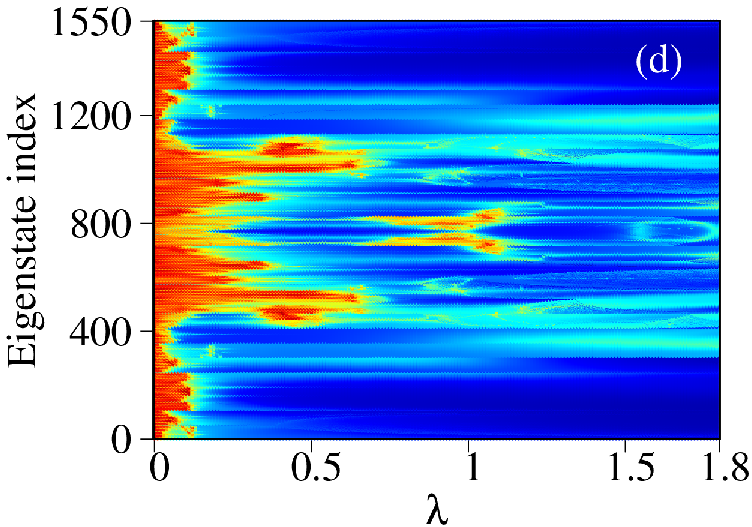}
\caption{(Color online). Density plots of the fractal dimension $D_n$. (a), (b) Energy spectrum as a function of modulation strength $\lambda$ for the Fibonacci and BM sequences, respectively. (c), (d) Density plots of the fractal dimension $D_n$ for individual eigenstates as functions of the eigenstate index and modulation strength $\lambda$ for the Fibonacci and BM sequences, respectively. The hopping parameters are fixed at $t_A = 1$ and $t_B = 2.5$, with system sizes $N = 1598$ (Fibonacci) and $N = 1550$ (BM). The colorbar indicates the values of $D_n$ for the $n$th eigenstate. The eigenstate indices are ordered according to the ascending values of their corresponding eigenvalues.}
 \label{fig:2} 
\end{figure}

To characterize the localization properties of one-dimensional chains with staggered AAH potentials, we begin by examining the fractal dimension $D_n$ of individual eigenstates. Figures~\ref{fig:2}(a) and (b) display density plots of $D_n$ across the energy spectrum as a function of the modulation strength $\lambda$ for the Fibonacci and BM sequences, respectively. In the colorbar scheme, red regions correspond to $D_n \rightarrow 1$, indicating extended states, blue regions correspond to $D_n \rightarrow 0$, signifying localized states, and gray areas represent intermediate or critical phases. The evolution of the energy spectrum with increasing $\lambda$ differs between the Fibonacci and BM cases due to the distinct forms of their incommensurate potentials, characterized by the parameter $b$ in Eq.~\ref{siteeng}. Owing to this incommensurability, the spectrum fragments into a fractal set of bands and gaps~\cite{hoffstad,kraus}.

To provide clearer insight into the localization behavior of the states, we further present density plots of $D_n$ as functions of the eigenstate index and modulation strength $\lambda$ for the Fibonacci and BM sequences in Figs.~\ref{fig:2}(c) and (d), respectively. For small $\lambda$, the quasiperiodic potential is relatively weak compared to the hopping amplitudes, resulting in predominantly extended states, as evidenced by the uniformly red regions across all eigenstates. As $\lambda$ increases, the system enters a critical regime in which the spatial inhomogeneity induced by the quasiperiodic potential begins to compete with kinetic delocalization. This interplay leads to the coexistence of extended and localized states within the same energy spectrum, giving rise to a single-particle mobility edge (SPME). 
The onset of this mixed phase occurs at lower $\lambda$ for the BM sequence than for the Fibonacci, indicating that the underlying 
substitution rule and the resulting local bond structures significantly influence the delocalization-localization transition.
Further increasing $\lambda$ leads to the full localization of all eigenstates (blueish regions), as the on-site disorder dominates the hopping.
However, quite interestingly, we notice that beyond a certain $\lambda$, delocalized states re-emerge, evident from red streaks in
both  Figs.~\ref{fig:2}(c) and (d). This reentrant localization-delocalization transition is rare and cannot occur in systems governed by monotonic disorder 
(e.g., random potentials), where localization is typically enhanced monotonically with increasing disorder. Instead, it reflects the nonlinear competition between two incommensurate modulations, the quasiperiodic potential and the bond alternation, leading to resonant tunneling across segments of the lattice. 

\begin{figure}[ht]
\centering \includegraphics[width=0.7\linewidth]{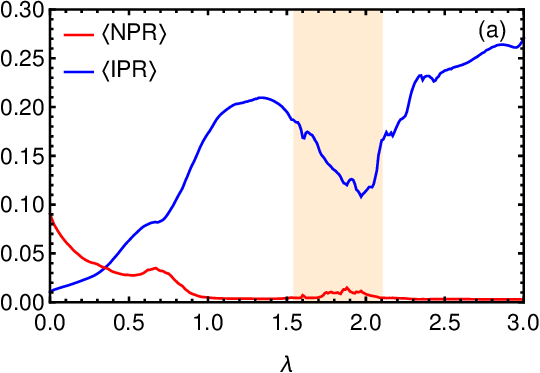}
\vskip 0.1 in
\includegraphics[width=0.7\linewidth]{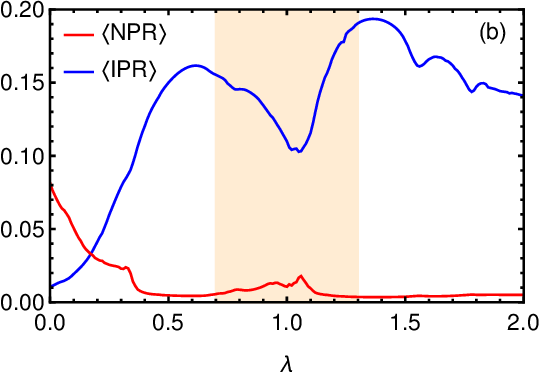}
\caption{(Color online). Average IPR and NPR for a selected subset of eigenstates as a function of $\lambda$. (a) Fibonacci and (b) BM sequences.  The subset of eigenstates are considered for the Fibonacci sequence between 28-72\% of the total states as depicted in Fig.~\ref{fig:2}(a) and that for the BM case is between 41-59\% of the total states as shown in Fig.~\ref{fig:2}(b). The system parameters are identical to those used in Fig.~\ref{fig:2}. The blue and red curves denote the results for the average IPR and NPR, respectively.}
\label{fig:3} 
\end{figure}

To substantiate this claim, we examine the average inverse participation ratio and the normalized participation ratio, shown 
in Fig.~\ref{fig:3}. The averaging for $\langle \text{IPR} \rangle$ and $\langle \text{NPR} \rangle$ is performed over a subset of eigenstates within the range of 28\% to 72\% of the total states for the Fibonacci sequence, and 41\% to 59\% for the BM case, as depicted in Fig.~\ref{fig:2}. Such choices of subsets ensure a more accurate and representative characterization of the localization properties in the system and have been consistently adopted throughout the paper. It is important to note that, in the RL region, only a certain fraction of states become delocalized. Therefore, if the entire set of eigenstates were considered, identifying the RL regions from the NPR and IPR plots would be difficult. Accordingly, for averaging the IPR and NPR, only the central $44\%$ of the spectrum in Fig.~\ref{fig:2}(c) is considered for the Fibonacci case, and the central $18\%$ of the spectrum in Fig.~\ref{fig:2}(d) for the BM case, corresponding to the energy windows within which the RL regions occur. For both sequences, the average NPR remains finite for low $\lambda$, confirming delocalized behavior. 
With increasing $\lambda$, the average NPR value starts to drop and the average IPR rises. Further, $\langle \text{NPR} \rangle$ becomes vanishingly small near $\lambda\sim 1$ for the Fibonacci sequence and $\lambda \sim 0.4$ for the BM case, reflecting a 
fully localized phase for both the sequences. Interestingly, with further increase in the modulation strength, the value of $\langle \text{NPR} \rangle$ becomes finite within the $\lambda$-range marked by the light-peach patch in both sequences, indicating that some previously localized states become extended again, indicating the emergence of reentrant behavior. This RL is observed for the Fibonacci sequence approximately within the modulation strength window $1.55 \lesssim \lambda \lesssim 2.1$, and for the BM case within $0.7 \lesssim \lambda \lesssim 1.3$. Physically, this suggests that constructive interference 
and long-range correlations, inherent to the quasiperiodic geometry, facilitate the formation of extended states even under 
significant potential modulation.

\begin{figure}[ht]
\centering \includegraphics[width=0.8\linewidth]{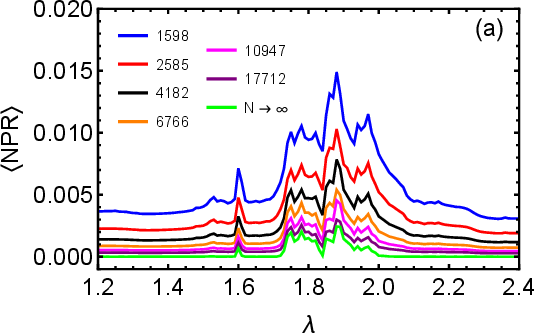}
\vspace{1em}
\includegraphics[width=0.8\linewidth]{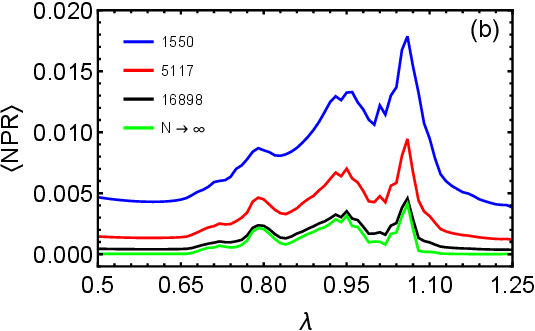}
\caption{(Color online). $\langle \text{NPR}\rangle$ as a function of $\lambda$ for the (a) Fibonacci and
(b) BM sequences. Various system sizes are considered to access the thermodynamic limit, $N\rightarrow\infty$. For the Fibonacci case, the considered system sizes are $N=1598, 2585, 4182, 6766, 10947, 17712$, and $N\rightarrow\infty$, and their corresponding results are represented by blue, red, black, orange, magenta, purple, and green, respectively. The system sizes chosen for the BM case are $N=1550, 5117, 16898$, and $N\rightarrow\infty$, denoted with blue, red, black, and green colors, respectively. All the other system parameters are the same as described in Fig.~\ref{fig:2}. The averaging scheme is the same as mentioned in Fig.~\ref{fig:3}.}
\label{fig:4} 
\end{figure}

To investigate the finite-size scaling behavior in the RL region, we compute $\langle \text{NPR} \rangle$ for various system sizes, as shown in Fig.~\ref{fig:4}. The averaging procedure is identical to that described in Fig.~\ref{fig:3}. 
The system sizes considered for the Fibonacci sequence are $N = 1598$, $2585$, $4182$, $6766$, $10947$, and $17712$, and for the BM sequence as $N = 1550$, $5117$, and $16898$. All other physical parameters, such as the hopping amplitudes $t_A$ and $t_B$, as well as the on-site potentials, are the same as in Fig.~\ref{fig:2}. Based on the computed data, we further extrapolate $\langle \text{NPR} \rangle$ in the thermodynamic limit, $N \rightarrow \infty$. Remarkably, $\langle \text{NPR} \rangle$ remains finite for both sequences within the reentrant region,  indicating the persistence of the RL phase as the system size increases and approaches the thermodynamic limit. This finite value of $\langle \text{NPR} \rangle$ confirms the robustness of the RL phase against finite-size effects. In contrast, outside the RL region, $\langle \text{NPR} \rangle$ vanishes, consistent with expectations for a localized phase.

To gain spatial insight into individual states, we examine local probability 
\begin{figure}[ht]
\includegraphics[width=0.241\textwidth]{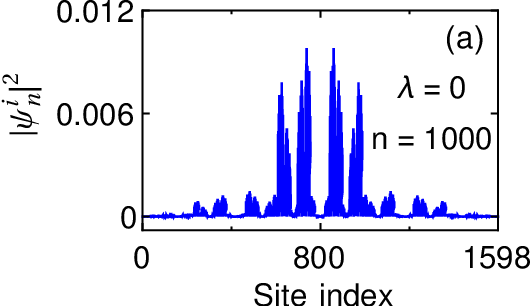}\hfill\includegraphics[width=0.241\textwidth]{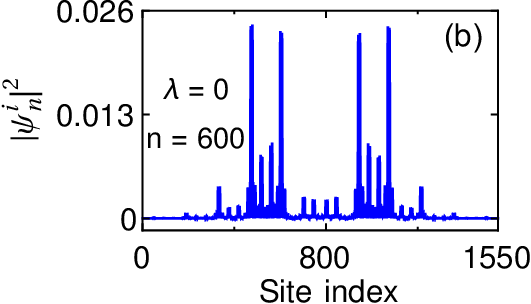}\vskip 0.1 in
\includegraphics[width=0.241\textwidth]{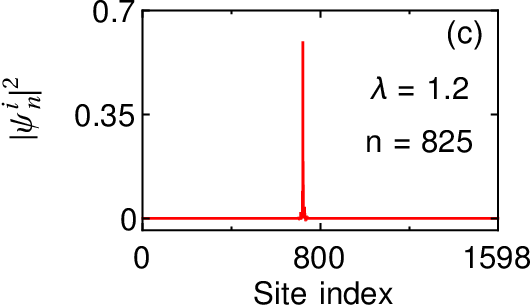}\hfill\includegraphics[width=0.241\textwidth]{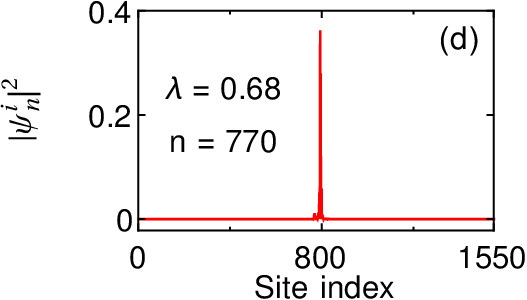}\vskip 0.1 in
\includegraphics[width=0.241\textwidth]{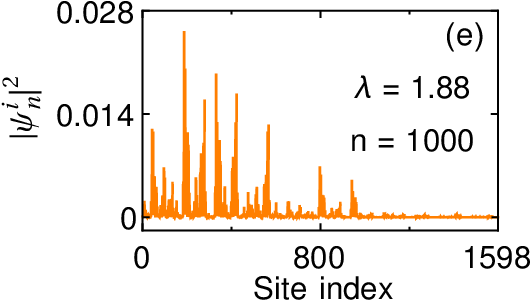}\hfill\includegraphics[width=0.241\textwidth]{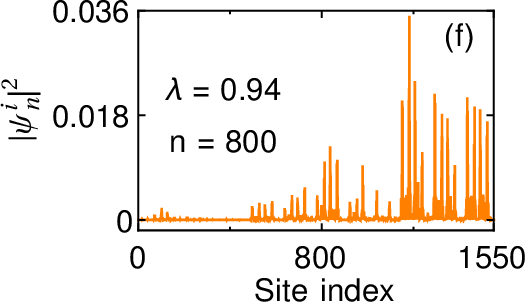}\vskip 0.1 in\includegraphics[width=0.241\textwidth]{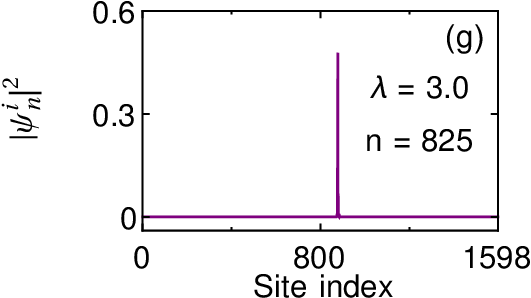}\hfill\includegraphics[width=0.241\textwidth]{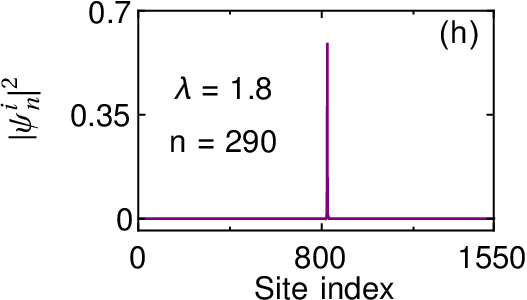}
\caption{(Color online).  Distribution of local probability amplitude $\lvert\psi _n^i \rvert^2$ as a function of site index $i$ for the $n$th eigenstate. Left (Right) column represents the Fibonacci (BM) sequence. 
For the Fibonacci sequence (a) $\lambda = 0$ and $n=1000$, (c) $\lambda = 1.2$ and $n = 825$, (e) $\lambda = 1.88$ a nd $n=1000$, and (g) $\lambda = 3.0$ and $n = 825$. For the BM case  (b) $\lambda = 0$ and $n=600$, (d) $\lambda = 0.68$ and $n = 770$, (f) $\lambda = 0.94$ and $n=800$, and (h) $\lambda = 1.8$ and $n = 290$. The system size and all the other physical parameters are identical to Fig.~\ref{fig:2} for the Fibonacci and BM cases.}
\label{fig:5}
\end{figure}
distributions $\lvert\psi_n^i\rvert^2$ ($n$ being the eigenstate index and $i$ the site index) across different regimes, as shown in Fig.~\ref{fig:5}. The left and right panels of Fig.~\ref{fig:5} depict the local probability distribution for the Fibonacci and BM sequences, respectively, at different modulation strengths. The number of sites are fixed as $N=1598$ for the Fibonacci case and 1550 for the BM case. All other system parameters are same as described in Fig.~\ref{fig:2}. 
At $\lambda=0$, the states are clearly extended as the probability amplitudes are very low across the entire chain as shown in Figs.~\ref{fig:5}(a) and (b) for the Fibonacci and BM sequences, respectively. At $\lambda=1.2$ for the Fibonacci chain, the wave function is sharply peaked (Fig.~\ref{fig:5}(c)) around a single site, confirming Anderson-like localization driven by the incommensurate potential. The same is also observed for the BM chain in Fig.~\ref{fig:5}(d) at $\lambda=0.68$. 
Intriguingly, at $\lambda = 1.88$, where the RL phase emerges, the wave functions once again spread across the system (Fig.~\ref{fig:5}(e)). Although the envelope is nonuniform but the amplitudes are very small, and thus the state clearly extends over the entire Fibonacci chain, indicating an extended phase. A similar behavior is observed for the BM chain at $\lambda = 0.94$, as shown in Fig.~\ref{fig:5}(f). At higher modulation strength ($\lambda = 3.0$), localization becomes significantly more pronounced in the Fibonacci case (Fig.~\ref{fig:5}(g)), reaffirming the eventual transition into a strongly localized phase. For the BM chain, this final localization transition occurs at $\lambda = 1.8$, as depicted in Fig.~\ref{fig:5}(h), beyond which no further reentrant behavior is observed within the given parameter setup.

So far, the hopping strengths $t_A$ and $t_B$ have been fixed at $1$ and $2.5\,\text{eV}$, respectively. However, to investigate how hopping correlations influence the localization properties, we evaluate the localization indicator $\eta$ over a broad region of the $t_B/t_A - \lambda $ parameter space. The quantity $\eta$ is defined as~\cite{li2020mobility}
\begin{equation}
\eta = \text{log}_{10}\left[\langle \text{IPR}\rangle\times \langle \text{NPR}\rangle\right].
\end{equation}
In the mixed-phase region, both $\langle \text{IPR} \rangle$ and $\langle \text{NPR} \rangle$ remain finite and of order $\mathcal{O}(1)$, yielding $\eta$ values in the range $-2 \leq \eta \leq -1$. In contrast, for a purely localized (extended) phase, $\langle \text{NPR} \rangle$ $\left( \langle \text{IPR} \rangle \right)$ scales as $N^{-1}$, while $\langle \text{IPR} \rangle$ $\left( \langle \text{NPR} \rangle \right)$ stays of order $\mathcal{O}(1)$. Consequently, for a purely localized or extended state, $\eta < -\log_{10} N$. For instance, when $N \approx 10^3$, one finds $\eta < -3$. This makes $\eta$ a reliable parameter for distinguishing a mixed phase from a completely localized or fully extended phase.

\begin{figure}[ht]
\centering \includegraphics[width=0.44\textwidth]{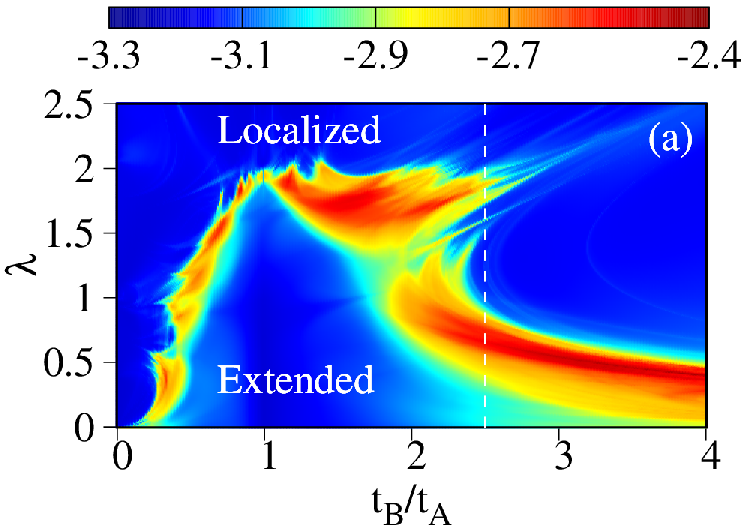}
\vskip 0.2 in
\includegraphics[width=0.44\textwidth]{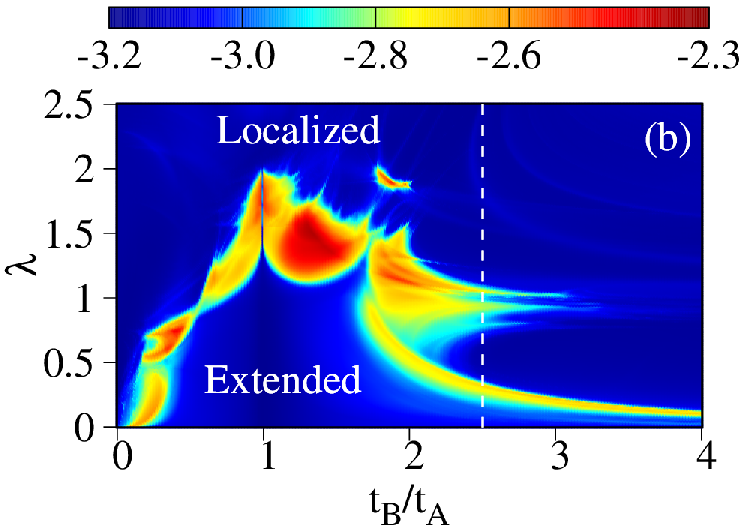}
\caption{(Color online). Density plots of $\eta$ in the $t_B/t_A-\lambda$ plane for (a) Fibonacci and (b) BM sequences. The color scale represents $\eta$ values, with dark blue and bluish regions denoting highly localized or fully extended phases, and red-yellow regions indicating mixed-phase zones. The white dashed line at $t_B/t_A = 2.5$ marks the parameter path used to track the RL transition. System sizes and all other physical parameters are identical to those in Fig.~\ref{fig:2}. The averaging scheme for $\langle \text{IPR} \rangle$ and $\langle \text{NPR} \rangle$ used to compute $\eta$ is the same as described in Fig.~\ref{fig:3}.}
\label{fig:6} 
\end{figure}

Figures~\ref{fig:6}(a) and (b) present the density plots of $\eta$ in the $t_B/t_A-\lambda$ plane for the Fibonacci and BM sequences, respectively. The system sizes and all other physical parameters are identical to those used in Fig.~\ref{fig:2}, and the averaging procedure for $\langle \text{IPR} \rangle$ and $\langle \text{NPR} \rangle$ is the same as described for Fig.~\ref{fig:3}. The colorbars indicate the $\eta$ values, where dark blue and bluish regions correspond to highly localized or fully extended states, while the red to yellowish regions represent mixed-phase zones. The classification of localized and extended regions is based on the combined analysis of $\langle \text{IPR} \rangle$ and $\langle \text{NPR} \rangle$. A vertical line at $t_B/t_A = 2.5$ is drawn to illustrate the parameter path associated with the RL transition. This parameter choice corresponds to the energy spectra shown in Fig.~\ref{fig:2}. 

The RL transition is evident in both the cases. The approximate $t_B/t_A$ window supporting RL is $2.3$-$3.0\,$eV. For a fixed $t_B/t_A = 2.5$, the RL transition occurs at higher $\lambda$ values in the Fibonacci case ($\lambda \approx 1.5$-$2$) compared to the BM case ($\lambda \approx 0.8$-$1.2$). Additional noteworthy features are also observed in the BM sequence. First, a small island of mixed phase appears near $\lambda \approx 2$ and $t_B/t_A \approx 2.0$, suggesting the possibility of an RL transition at a different $t_B/t_A$ value than the main RL region. Second, near $t_B/t_A \approx 3.0$, multiple RL transitions are observed. For example, along $t_B/t_A \approx 2.8$, three distinct sky-blue patches (corresponding to mixed-phase zones) appear, suggesting the occurrence of three RL transitions.

Overall, our results demonstrate that the presence of two distinct types of off-diagonal disorder, combined with staggered AAH on-site modulation, leads to an unconventional two-step localization-delocalization-localization (LDL) transition, separated by two distinct SPMEs. This behavior originates from nontrivial correlations between site energies and hopping amplitudes, which arise directly from the underlying quasiperiodic sequence. These correlations are particularly evident in the choice of the incommensurability parameter $b$ in the on-site AAH potential. The RL phenomenon emerges in this work for the Fibonacci chain with $b = (1+\sqrt{5})/2$ and the BM chain with $b = (3+\sqrt{13})/2$, corresponding to the asymptotic ratios of the number of building blocks $t_A$ to $t_B$ in the respective sequences.

The two different choices of $b$ can be justified as follows. In standard AAH models, $b$ is typically taken as a fixed irrational number (e.g., the inverse golden mean), independent of the hopping pattern. In our case, the hopping amplitudes follow a deterministic substitution sequence, and we set $b$ equal to the asymptotic $t_A/t_B$ ratio of that sequence. This choice aligns the on-site potential with the hopping pattern, producing a correlated quasiperiodic structure rather than two statistically unrelated modulations. Such commensurability enhances constructive interference between diagonal and off-diagonal quasiperiodicities, enabling the stable extended regimes that give rise to RL. In contrast, if $b$ is chosen as a generic irrational number, the two modulations become statistically incommensurate, leading to monotonic localization without intermediate extended phases. We have verified this by testing several other irrational $b$ values for both Fibonacci and BM hoppings. We observe that RL occurs robustly not only when $b$ matches the asymptotic symbol ratio of the sequence (i.e., $b = (1+\sqrt{5})/2$ for Fibonacci and $b = (3+\sqrt{13})/2$ for BM), but also for their inverses, as well as for twice these values and twice their inverses. This suggests that similar kinds of $b$, closely related to the fundamental inflation parameters, may also produce RL, highlighting the importance of correlated quasiperiodicity in the observed two-step LDL transitions.

Moreover, the Fibonacci chain exhibits this two-step transition more prominently and across a broader disorder range, while the BM chain shows qualitatively similar behavior but over narrower intervals. This contrast suggests that the robustness of extended states is closely linked to the self-similarity and substitution rules defining the sequence. We have also examined other substitutional sequences, such as Thue-Morse, Copper mean, and those listed in Table~1 of Ref.~\cite{guopre2007}, to introduce correlated off-diagonal disorder. However, within the parameter regime considered here, we did not observe RL in these cases. The corresponding results are therefore omitted for clarity and conciseness, although RL may still emerge for these sequences under different conditions, making this a promising direction for future work.

From a broader perspective, a universal feature across all known RL studies, including the present work, is the presence of a staggered quasiperiodic on-site potential, i.e., alternate-site energies modulated in an AAH-type manner. The thermodynamic-limit behavior of global localization measures, such as finite NPR values within RL windows~\cite{roy2021reentrant, wangprb2023, liarxiv2023, ganguly2023spin}, is also consistent across systems. Moreover, we share with prior works~\cite{liarxiv2023, ganguly2023spin} the observation of multiple RL transitions, as well as a parameter-space organization in which extended and localized phases are interspersed with mixed-phase regions.

The main distinction lies in the hopping structure. Whereas most earlier studies achieve RL through periodic hopping modulation, typically dimerization, or via extended-range couplings, our models realize RL entirely through correlated quasiperiodic hopping, without introducing such periodicity. This approach changes the microscopic origin of the modulation while retaining the staggered quasiperiodic potential, thereby providing a new minimal route to RL. Our $\eta$-maps reveal broad mixed-phase bands and smaller mixed-phase islands, particularly in the BM sequence, producing cascade-like fragmentation reminiscent of that reported in~\cite{goblot2020emergence, zhai2021cascade, padhanprb2022}, yet arising here from a fundamentally different hopping mechanism.

\section{Closing Remarks} 

In this work, we have demonstrated the emergence of the RL phenomenon in one-dimensional quasiperiodic systems, specifically in the Fibonacci and BM hopping models subjected to a staggered AAH on-site potential. This finding introduces a new class of minimal models that support RL purely through off-diagonal correlated disorder in the presence of staggered quasiperiodic modulation.

By analyzing the fractal dimensions of eigenstates along with the behavior of the IPR and NPR across the energy spectrum, we identified critical regions characterized by SPMEs. These regions separate localized and extended eigenstates and are bounded by two critical modulation strengths, giving rise to an RL phase. Within this phase, certain eigenstates that were initially localized become extended again upon increasing modulation strength, an unconventional phenomenon not typically observed in traditional localization scenarios. Finite-size scaling of NPR across various system sizes confirms the robustness of the RL phase in the thermodynamic limit, ruling out finite-size artifacts. Local probability amplitude distributions further reveal spatial eigenstate characteristics across modulation strengths, highlighting transitions between extended, localized, and re-extended states. The $\eta$-maps have revealed that RL transitions occur in both Fibonacci and BM sequences, with the RL window spanning a finite range of $t_B/t_A$. For a representative value within this range, the RL transition has appeared at higher $\lambda$ values in the Fibonacci case than in the BM case. The BM sequence has further exhibited distinctive features, including a small isolated mixed-phase region suggesting a possible secondary RL window, as well as multiple RL transitions indicative of a more fragmented phase structure. Our results have shown that RL can arise for a range of quasiperiodic modulation parameters closely related to the fundamental inflation ratios, emphasizing the crucial role of correlated quasiperiodicity in driving such transitions.

Given the growing interest in quasiperiodic systems and the successful experimental realization of AAH-type models in photonic lattices,
ultracold atomic systems, and waveguide arrays~\cite{lahini2009observation, kraus2012topological, wang2020localization, roati2008anderson, christodoulides2003discretizing}, we believe that the reentrant localization phenomena predicted in this study are experimentally 
accessible. Our findings may open new avenues for engineering localization-delocalization transitions in synthetic quantum systems and 
could potentially inform the design of novel quantum devices based on tunable transport properties. 

\section*{Acknowldedgments}

SK acknowledges financial support from the University Grants Commission (UGC), India (NTA Ref. No. 211610108524), through a research fellowship.

\end{document}